# Meta-analyses of dietary exposures must consider energy adjustment: recommendations from a meta-scientific review

Natalia Ortega[1,2*], Peter WG Tennant[3,4], Darren C Greenwood[3,4], Octavio Pano[2,5], Christina C Dahm[6], Russell J de Souza[7,8,9], Daniel B Ibsen[6,10], Conor J MacDonald[11], Deirdre K Tobias[2,5,12], Georgia D Tomova[13]

**Affiliations**

[1] Institute of Primary Health Care (BIHAM), University of Bern, Bern, Switzerland

[2] Division of Preventive Medicine, Department of Medicine, Brigham and Women's Hospital, Boston MA, USA

[3] Leeds Institute for Data Analytics, University of Leeds, Leeds, UK

[4] School of Medicine, University of Leeds, Leeds, UK

[5] Harvard Medical School, Boston MA, USA

[6] Department of Public Health, Aarhus University, Bartholins Allé 2, 8000 Aarhus C, Denmark

[7] Department of Health Research Methods, Evidence, and Impact, McMaster University, Hamilton, ON L8R 2K3, Canada

[8] Mary Heersink School of Global Health and Social Medicine, McMaster University, Hamilton, ON, Canada

[9] Population Health Research Institute, Hamilton Health Sciences Corporation, Hamilton ON, Canada

[10] Steno Diabetes Center Aarhus, Aarhus University Hospital, Palle Juul-Jensens Boulevard 11, 8200 Aarhus N, Denmark

[11] Unit of Epidemiology, Institute of Environmental Medicine, Karolinska Institutet, Solna, Sweden

[12] Nutrition Department, Harvard TH Chan School of Public Health, Boston, MA

[13] Centre for Longitudinal Studies, University College London, UK

**Corresponding author:** Natalia Ortega, nherrero@bwh.harvard.edu, 900 Commonwealth Av 3rd Floor Boston, MA, US

**Word count:** 4,902

## Abstract

**Background:**

In observational studies of dietary exposures, the energy adjustment strategy has a critical impact on the effect being estimated. Adjusting for total energy intake or expressing the exposure as a percentage of total energy (i.e., isocaloric model), leads to a substitution effect being estimated. This has important implications for the interpretation of individual studies and meta-analyses. The energy adjustment strategy should be considered in meta-analyses to ensure comparable effects are pooled.

**Aim:**

This meta-scientific review aimed to investigate the extent to which meta-analyses of dietary exposures are pooling incomparable effects, by reviewing the energy adjustment strategies in an illustrative sample of eight meta-analyses of diet and cardiovascular disease risk and their constituent primary studies.

**Methods:**

We identified all meta-analyses examining the relationship between 1) saturated fat (an example macronutrient) and cardiovascular disease; and 2) fish (an example food) and cardiovascular disease published through December 3rd 2023 by searching Medline, Embase, the Cochrane Database of Systematic Reviews, and Web of Science. The two most recent and two most cited reviews for each exposure (saturated fat, fish) were examined, along with all constituent studies. Information on the study aims, the effects being targeted by different models, and result interpretations was extracted and summarized.

**Results:**

The eight meta-analyses summarised results from 82 original research manuscripts (34 saturated fat, 48 fish) reporting on 144 unique models (63 saturated fat, 81 fish). Only one meta-analysis explicitly considered the energy adjustment strategy of the primary studies in its evidence synthesis to determine eligibility for a substitution subgroup analysis. None of the meta-analyses acknowledged that they were pooling estimates for different effects. 82% (n=119/144) of the models from the primary studies —94% (n=59/63) of the models in studies of saturated fat and 74% (n=60/81) of the models in studies of fish— were implicitly estimating substitution effects but this was not explicitly stated in most study aims, interpretation, or conclusions.

**Discussion:**

Our meta-scientific review of eight meta-analyses evaluating diet and cardiovascular disease indicated that the energy adjustment strategies of the primary studies were sparsely considered in the synthesis and interpretation of evidence. Consequently, the pooled estimates reflect ill-defined quantities with unclear interpretations. We offer recommendations to improve the conduct of future meta-analyses to improve the quality of evidence that informs nutritional and public health recommendations.

**Funding:** NO was awarded a "UniBE Short Travel Grants for (Post)Docs" and a Postdoc.Mobility grant by the Swiss National Science Foundation (n 230635).
**Registration:** This review was pre-registered on the Open Science Framework (https://doi.org/10.17605/OSF.IO/NMB5Y).

**Introduction**

Dietary recommendations are a cornerstone of public health interventions. Because of the challenge of studying the long-term effects of diet using experimental approaches, nutrition recommendations for overall health promotion are typically based on findings from large-scale observational studies and meta-analyses. However, estimating causal effects from observational data is notoriously challenging. Recent guidelines therefore advise that observational studies should follow a transparent framework of pre-specifying the target causal questions and clearly outlining all the relevant assumptions on which the estimates rely (1).

Causal questions in nutritional epidemiology primarily focus on studying the effects of foods (e.g., fish), micronutrients (e.g., vitamin D), or macronutrients (e.g., saturated fats). Two broad types of causal effect are typically of interest: (i) the 'addition effect' of either increasing or decreasing the intake of a dietary component (such as a food or macronutrient) while keeping the rest of the diet unchanged (also known as the total compositional effect (2,3)) and (ii) the isocaloric 'substitution effect' of increasing the intake of one dietary component and decreasing one or more other defined dietary components, while maintaining the same total energy intake (also known as the relative compositional effect (2,3)). Different statistical approaches are used to estimate these two types of effects with observational data, which are explained in Box 2 (4–8).

Although several articles have explained how different energy adjustment strategies can lead to different effects being estimated, these important concepts are still not universally recognised or acknowledged by applied researchers (4–8). More commonly, total energy intake and other dietary components are considered as 'potential confounders' for the diet-disease effect of interest, serving as the rationale for investigators to consider various statistical adjustments without necessarily explicitly detailing how this how these modelling decisions might change the quantity being estimated, as explained in **Figure 1**.

**[Figure 1 here]**

Although energy-adjustment modelling choices originate in the primary studies, there are also significant implications of these approaches for the conduct, interpretation, and validity of meta-analyses (9). Meta-analyses pool the results of primary studies reporting on the same research question. However, meta-analyses of dietary exposures that do not take account of the role of the primary study's energy adjustment strategies may inappropriately pool estimates for different quantities, impairing the interpretability and relevance of the results. To investigate the extent that this may be an issue in practice, we conducted a meta-scientific review of a sample of meta-analyses and their constituent primary studies to examine how and whether studies consider the role of the energy adjustment strategy in the effects being estimated, and the degree to which the reported estimates correspond with the primary study aims and interpretations. We consider the examples of saturated fats (as a macronutrient exposure) and fish (as a food exposure) in relation to cardiovascular disease incidence (as the outcome). We present a set of simple recommendations to improve the conduct, presentation, and interpretation of future meta-analyses of nutritional exposures.

**Methods**

*Overview*

This review sought to extract, examine, and summarise relevant information on the study aims, language use, approach to energy adjustment, reporting of results, and interpretations from a sample of eight meta-analyses of diet and cardiovascular disease (four focussed on saturated fats and four focussed on fish) and their 82 relevant constituent studies. We were particularly interested in the extent to which the meta-analyses acknowledged and accounted for the fact that different adjustment strategies can lead to different effects being estimated. We were also interested in the justification for different energy adjustment strategies in the primary studies, and whether these aligned with the study aims and inferences. We chose to examine two different dietary exposures, one macronutrient (saturated fats) and one food (fish) because of important differences in how each can be modelled with respect to total energy adjustment.

This review was pre-registered on the Open Science Framework (10) and adheres to the PRISMA Reporting Guidelines for meta-analyses (11).

*Data sources*

We conducted systematic searches to identify all meta-analyses examining the effects of fish and saturated fats on cardiovascular disease incidence. We searched Medline (Ovid), Embase (Elsevier), Cochrane Database of Systematic Reviews (Wiley), and Web of Science (Clarivate) on 8 December 2023, without language restrictions. The syntax for each of the searches are provided in Section 1 of **Supplementary File**.

*Study selection*

We included meta-analyses that reported effect measures combining observational studies evaluating the relationship between dietary intake of saturated fat or fish intake with cardiovascular disease incidence in populations initially free of cardiovascular disease. We excluded studies reporting estimates only for subtypes of saturated fats or fish species, cardiovascular disease mortality, or populations exclusively with pre-existing cardiovascular disease (i.e., secondary prevention).

Two reviewers (NO and OP) independently screened titles, abstracts, and full texts of potentially eligible studies. Discrepancies were resolved by a third independent reviewer (GDT). After identifying all relevant meta-analyses, we selected the two most cited and the two most recent studies for each exposure. This allowed us to examine both how well the most influential meta-analyses in the field have been conducted and reflect on whether the conduct may have changed with more contemporary methodological practices in more recent publications.

*Data extraction*

Separate data extraction forms were developed for the meta-analyses and their constituent primary observational studies (see Sections 2 and 3 in **Supplementary File)**. We excluded any study that was not observational and that was not a full research article (e.g., conference abstracts, commentaries, letters). Primary studies were randomly assigned to pairs of co-authors who independently extracted data on the reporting of study aims and the target causal effects of interest, exposure specification, fitted models (number and justification for the adjustment sets), energy adjustment strategies, reporting of the results, discussion of the energy adjustment strategy and reporting and interpretation of the results in the conclusion. For primary studies that examined the exposure in multiple forms (e.g., categorical and continuous; kcal/d and % total calories), we extracted the first two reported versions, unless described as sensitivity analyses. We did not extract unadjusted models, unless no adjusted models were reported. We considered all adjusted models reported in primary analyses with different dietary variables. Where multiple models contained the same dietary variables, we extracted data for the most comprehensively adjusted model, unless it was a sensitivity analysis or included variables described as mediators. Where it was unclear whether a model was a primary or sensitivity analysis, it was included. We also extracted the authors' interpretation (addition or substitution) of the results. Discrepancies were resolved by a third reviewer (one of NO, PWGT and GDT). A complete list of all included articles is provided in Sections 4 and 5 of **Supplementary File**.

*Data synthesis and analysis*

We report descriptive findings (counts and proportions) for the main characteristics of the primary studies and meta-analyses. We first considered whether a target causal effect of interest was stated in the aim. Next, we examined the adjustment set(s) to determine which effects were actually targeted by the primary studies, classifying these into broadly effects of addition or substitution. Addition effects were identified when a model included the exposure in mass or caloric units and did not adjust for total energy intake, including where adjustment was made for remaining energy (i.e., the energy partition model) or all other components (i.e., comprehensive energy partition model) (**Figure 1**). Isocaloric substitution effects were identified when a model adjusted for total energy intake or used an energy-adjusted residual as the dietary exposure, whether intentionally (i.e. the standard model, residual model, or leave-one-out-model) or otherwise (e.g., when total energy is adjusted as a 'potential confounder'), expressed the exposure as a percentage of total energy intake (i.e., the nutrient density model), or both (i.e. the multivariable nutrient density model) (4–8) (**Figure 1**). The specific substitution being targeted depended on several factors, including which modelling approach was used, how total energy intake was adjusted for, which other dietary variables were also adjusted for, and whether the exposure and any other components were in mass or caloric units. We compared our interpretation of whether the effects being targeted represented addition or substitution to the authors interpretation.

**Results**

*Literature flow*

The searches returned 2,929 unique records. After full text screening, 24 meta-analyses met the inclusion and exclusion criteria, with 8 reporting on saturated fats and 16 on fish. The final sample included 1) the two most cited meta-analyses studying saturated fats and cardiovascular disease incidence, with 2,010 and 1,815 citations each (as of 4 January 2024) (12,13), 2) the two most cited meta-analyses studying fish intake and cardiovascular disease incidence, with 767 and 519 citations each (as of 4 January 2024) (14,15), 3) the two most recent studying saturated fats and cardiovascular disease incidence (as of 8 December 2023 (16,17), and 4) the two most recent (8 December 2023) studying fish intake and cardiovascular disease incidence, published in 2021 and 2023 (18,19) (**Supplementary Figure**). Section 4 provides full details of the 8 meta-analyses and Section 5 provides full details of their 82 primary studies (**Supplementary File**).

*Meta-analyses*

None of the eight meta-analyses had explicit causal aims or stated a target causal effect of interest. Five aimed to estimate "association(s)" and 3 aimed to estimate "relationship(s)" between the dietary exposures and cardiovascular disease outcomes (see **Table 1**).

Only one meta-analysis (13) considered the energy adjustment strategy as rationale for deciding which model results to extract and meta-analyse from each primary study for subgroup analyses of "substitution models", of which only one was successfully conducted (i.e., substitution of saturated fat for carbohydrates). Of the remaining seven meta-analyses, one preferentially extracted crude (unadjusted) associations, four extracted and analysed results from models declared the "most adjusted", one selected estimates from models based on their exclusion of any potential mediators, and one provided no information for which estimates were selected. None of the meta-analyses considered unit homogeneity of the exposure was in mass or caloric units.

Even though it was not possible to determine whether the authors of the meta-analyses were interested in evaluating addition or substitution effects of the dietary exposures on cardiovascular disease from their aims or model selection, two meta-analyses nevertheless described their findings using broadly addition language, four used language that was ambiguous, one used a mix of addition and ambiguous language (depending on whether the exposure was treated as continuous or categorical, respectively), and one used ambiguous language for the main analysis and substitution language for the substitution sub-analysis.

Three out of eight meta-analyses acknowledged in the methods or discussion that either adjustment for total energy intake (12,13) or coding the dietary exposure as a percentage of total energy (16) can introduce a substitution interpretation, and two recognised that adjustment (or lack thereof) for total energy intake may have contributed to heterogeneity in their meta-analysis, including one (19) that explicitly examined whether adjustment for total energy intake introduced between-study heterogeneity. Despite this, none of the meta-analyses discussed the possibility that they might have combined addition and substitution effects.

The eight meta-analyses included 84 unique primary studies, two of which we excluded from further consideration (one for being a randomised controlled trial and one for being a non-peer reviewed conference abstract). Of the remaining 82 primary studies, 34 examined saturated fats and 48 examined fish.

*Primary studies*

The primary studies are summarized in **Table 1**. None of the primary studies stated explicitly causal aims or a target causal effect of the dietary exposures on cardiovascular disease risk, although 14 (17%) used indirect causal language (e.g., "increasing the risk" or "influencing"). The majority described objectives to estimate "association(s)" (n=42, 51%) or "relationship(s)" (n=24, 29%). Only one study explicitly described a substitutive effect in their aim ("*we aimed to investigate the predicted effects of isocaloric substitutions of carbohydrates for fats (20)*").

From the 82 studies, we identified 144 unique adjusted statistical models, analysing variations of the dietary exposure and/or unique sets of dietary adjustment variables. Of the 63 saturated fat models, 52% (n=33/63) analysed saturated fat as percentage of total calories and 48% (n=30/63) as grams/day or grams/week. Of the 81 models for fish intake, none examined intake as a percentage of total calories, 42% (n=34/81) expressed fish in grams/day or grams/week, 35% (n=28/81) in servings or portions, and 23% (n=19/81) in frequencies (times/d, often, never).

Only twenty-two (15%) models provided a justification for their energy adjustment strategy, all which were models of saturated fat intake. Fourteen of these justified adjusting for total energy intake because it introduced an "isocaloric substitution" effect and eight justified the adjustment on the grounds that total energy correlates with total fats and is therefore a potential confounder. In 122 (85%) other models [41 (65%) saturated fat models and 81 (100%) fish models], the rationale for adjusting for total energy intake was not given, despite it being listed as a covariate as a 'potential confounder'. Only a quarter of these models (n=36, 25%) named an established energy adjustment method in their approach: 17 (12%) were described as residual models, 10 (7%) as nutrient density models, and six (4%) as multivariable nutrient density models. The remaining 108 (75%) models were not described using any specific names.

Based on their exposure and energy adjustment approaches, we classified 119 (83%) models as implicitly modelling dietary substitutions, including 59 (94%) saturated fat models and 60 (74%) fish models. We determined that 25 (17%) models were implicitly addition models, including 4 (6%) saturated fat models and 21 (26%) fish models. Among the 119 substitution models, 70 (59%) adjusted for total energy as a covariate, including 16 (27%) saturated fat models and 54 (90%) fish models, 8 (7%) were unadjusted nutrient density models, including 7 (12%) saturated fat models and one (2%) fish model, and 41 (34%) were multivariable nutrient density models, including 36 (61%) saturated fat models and five (8%) fish models. Of the 111 models that adjusted for total energy as a covariate, nine (8%) categorized the variable by quantiles of intake and the rest modelled it as a continuous variable.

The 59 saturated fat substitution models additionally adjusted for alcohol (n=28,47%), total protein (n=20, 34%), fiber (n=17, 29%), vegetables (n=13, 22%), dietary cholesterol (n=8, 14%), total carbohydrates (n=6, 10%), fruits (n=10, 17%), and/or polyunsaturated fats (n=5, 8%). The 60 fish substitution models most frequently adjusted for alcohol (n=30, 50%), fruits and vegetables (n=24, 40%), fibre (n=14, 23%), saturated fats (n= 14, 23%), polyunsaturated fats (n=8, 13%), red meat (n=6, 10%), carbohydrates (n=5, 8%) and dietary scores excluding fish (n=4), Only 13 (10%) models adjusted for all dietary variables in the same units, while 99 (83%) analysed one or more variables in different units, and seven (6%) did not adjust for any other dietary variable besides total energy (**Table 1**).

In the discussion, only 16 (13%) of the estimates from models that we identified as targeting substitution effects were actually recognised as substitutions, 7 (6%) were interpreted as being broadly about addition, and 96 (81%) were either interpreted with ambiguous language (e.g., "*the HR was 1.6 for the association between saturated fats and stroke*") or in terms of statistical significance only (i.e. *the estimate was not statistically significant*). Where explicitly acknowledged, substitution effects were described in various ways, such as "*the association of saturated fats in exchange for the average combination of types of fat other than saturated fats (i.e. mono and polyunsaturated fats* (21))" or "*when compared with 1 serving per day of red meat [...], 1 serving per day of fish was associated with a 24% (95% CI, 6% to 39%) lower risk*" (22).

In the models that we identified as targeting addition effects, no adjustment was made for either total or partitioned remaining energy intake, and none provided a justification for the lack of adjustment. In authors' discussion and interpretation of results, all 25 addition models were either described using ambiguous language or in terms of statistical significance only, such as "*There were no significant associations between SFA intake and coronary events* (23)" or "*Intake of total fish products was also associated with a significant 15% (95% CI: 25, 2%) reduction in total stroke death* (24)".

## Discussion

### *Summary of findings*

Our meta-scientific review explored whether meta-analyses of dietary exposures consider the impact on the estimates' interpretation of the analytic decisions for total energy adjustment in the primary studies. Among the eight meta-analyses that we selected to examine, only one, the most cited though oldest, considered the impact of the energy adjustment strategy of the primary studies. This meta-analysis investigating saturated fat intake and cardiovascular disease risk sought to conduct a secondary meta-analysis specifically among substitution models whereby saturated fat was contrasted with isocaloric intakes of polyunsaturated fat or carbohydrates; however, the authors' aim was not achieved due to the lack of original research articles publishing results from such substitution models. (13). None of the meta-analyses explicitly acknowledged that their main summary effects pooled estimates for different effects, although there was some recognition that different adjustment approaches may have contributed heterogeneity (19).

Our assessment of the primary studies indicated that for both saturated fat and fish, most of the research inherently conducted and reported results from models estimating substitution effects, but this was rarely evident from the authors' stated aims, presentation of results, or conclusions. The reason these models reflect substitution effects was most often due to the adjustment for total energy intake, which was routinely included as a covariate without explicit justification. There were, however, some differences in the reporting between the studies analysing saturated fat vs. fish. None of the fish studies provided a justification for their energy adjustment strategy, whereas around a quarter of the saturated fat studies justified adjusting for total energy on the grounds they were interested in an isocaloric substitution. For both saturated fat and fish, there was considerable variation between studies in terms of the other dietary variables included in the models, which will also impact the interpretation of the estimate. This means that although the primary studies were reporting on different causal effects, this was largely ignored in the systematic reviews that pooled one of them all into the meta-analyses, without using any criteria related to the model that estimated the effect of interest. Even within th(1e same study, it was common for authors to report estimates from different models, ultimately reflecting different causal effects. Only a small fraction of the substitution models adjusted for all dietary variables in the same units, which is essential for interpretability of the substitution estimates in compositional data theory (7). A few models adjusted for total energy intake as a categorical variable, which only partially adjusts for energy intake, producing uninterpretable partial substitution effects.

Overall, all of the meta-analyses in our illustrative assessment sample of highly cited or recently published studies overlooked the impacts of energy adjustment strategies among their primary studies when conducting their meta-analyses, and hence pooled estimates relating to very different research questions.

### *Analysis*

Despite previous literature on the importance of the energy adjustment strategy (4–8), the issue received limited attention within the eight meta-analyses that we examined. This corresponds with the attention given by the primary studies themselves. Lack of awareness, resource limitations, disciplinary conventions, and the limits of existing guidance may all explain why the energy adjustment strategy is typically overlooked. It is unfortunate that the justification of the energy-adjustment strategy is also not a feature of STROBE-nut (25), the natural reporting guideline for observational studies of dietary outcomes. Three of the meta-analyses we examined (12,14,16) used the Grading of Recommendations Assessment, Development and Evaluation (GRADE) criteria to evaluate the quality of the primary studies and another three (17–19) used the Newcastle-Ottawa Scale to assess risk of bias (26,27). The Newcastle-Ottawa Scale considers the appropriateness of the adjustment set for the target question under the comparability domain. Nutrition meta-analyses could specify the energy adjustment strategy (e.g., selection of studies that model an isocaloric substitution by adjusting for total energy intake) and the dietary components being substituted by the exposure the in the primary studies. Similarly, GRADE includes the indirectness domain that should be used to judge the degree that a study or model is targeting the effect of interest., and should be rated as low if the study fails to define the energy adjustment strategy according to the research question of interest. We encourage the explicit consideration of the energy adjustment strategy, also in other risk of bias tools such as ROBINS-I or Petersen's confounding matrix, because they are not specifically designed for nutrition epidemiology studies but they are flexible to accommodate the characteristics of nutrition meta-analyses (28,29).

Although our review did not aim to critically appraise the studies that we examined, we encountered some common methodological issues. Chief among these, for both the primary studies and the meta-analyses, was the lack of a clear aim. No meta-analysis or primary study declared a target causal effect of interest. Unclear aims impede proper consideration of the methods and assumptions required to estimate causal effects (30). Meta-analyses with loosely defined aims are also at risk of including and pooling different quantities (i.e., relating to answering different research questions), leading to findings that are difficult to interpret and may have lower relevance for dietary guidelines and public health interventions.

Considerable heterogeneity is likely to have been introduced by the differences in the effects being estimated by the primary studies and subsequently pooled by the meta-analyses. For example, with fish, the effects being pooled included the effect of adding fish, the substitution effect increasing fish instead of a mix of all other dietary components except alcohol (31,32), and the substitution effect of increasing fish instead of other sources of protein (meat, eggs and pulses) (33). The variation in the effects being targeted will have been further exacerbated by the widespread use of 'mixed-unit' models, i.e. where some dietary variables are expressed in portions or mass units while others (especially total energy intake) are expressed in calories (7). The resulting models target obscure effects that may diverge heavily from more well-defined substitution effects (5,7). We similarly saw a handful of studies adjusting for total energy intake in categorical form which would also target obscure effects that include a mix of addition and substitution elements. Taken together, this makes it difficult to interpret the results of the eight meta-analyses examined as they each pooled a range of different, often poorly defined, research questions.

**Limitations**

Our review has some important limitations. First, only an illustrative sample of eight meta-analyses from two nutrition research areas was selected. The results may not generalize to other dietary exposures and outcomes. The clear differences in practice between studies of saturated fats and studies of fish does suggest that sub-disciplinary variations might exist across the broader dietary research literature (34). Although the sampling approach prioritized recent meta-analyses, we found no peer-reviewed meta-analyses of observational studies of fat and fish that explicitly aimed to estimate substitution effects. However, novel meta-analyses that explicitly seek to estimate substitution effects have begun appearing across the literature (35–39). These address some but not all the issues identified and discussed herein and may be worth examining as examples of better practice, though the methods are still under development.

Second, to maximize clarity, we focussed our review on the energy adjustment strategy of the meta-analyses and their constituent studies. We did not consider the many other important aspects that contribute to the specific effect being estimated, such as the total covariate adjustment set, causal contrast of choice, target population, exposure and outcome definitions, magnitude or dosage of the exposure, timing of exposure and start and end of follow-up (44). All of these should be considered within the definition of the target causal effect of interest. However, most of these are not unique to the context of dietary exposures and therefore general guidelines for them are already available (40,41).

**Recommendations for researchers conducting a meta-analysis of dietary exposures**

Where possible, individual participant data meta-analyses can potentially avoid the problems discussed in this review by ensuring that all contributing datasets are analysed consistently to target the same effects. For those conducting conventional meta-analyses of dietary exposures without access to the individual participant data, we offer the following recommendations (also available as a checklist in Table 2). These are intended to complement, not replace, existing reporting guidance, quality appraisal tools and risk of bias tools for use in meta-analyses, such as PRISMA (11), MOOSE (42), GRADE (27).

**1) Clearly formulate the meta-analysis aims and target causal effects of interest**

- Clearly establish the study aims before conducting the literature search, and determine whether you are interested in addition effects, substitution effects, or both. Whether the interest lies in an addition or substitution effect may depend on how the dietary components are consumed in your target population, and whether you are more interested in evaluating a potential dietary intervention or studying a biological mechanism. Addition effects may be of more interest in foods or nutrients consumed in isolation or considered discretionary rather than essential, such as sugar-sweetened beverages and snacks. Substitution effects may be of more interest in foods or nutrients that are typically consumed as a core part of the diet and/or in combination with other foods or nutrients, such as meat, vegetables, or carbohydrates. If the interest lies in a substitution effect, consider which substitutions are specifically of interest, e.g., fish instead of any other food, fish instead of meat, or fish instead of white meat, each of which may be very different.

- Clearly formulate the study aims into target causal effects of interest using a recognised framework such as PICO(T) (43), providing clear descriptions of the target population, intervention, comparator, outcome, and timeframe of study. For example, "we aim to estimate the substitution effect of replacing 200 g of meat per week with 200 g of fish on the five-year risk of incident cardiovascular disease in non-vegetarian Southern-European populations".

**2) For each target causal effect of interest, identify all appropriate energy adjustment strategies and ideal variables for adjustment**
- All models should be clearly identified as estimating addition or substitution effects and specified in the protocol. Addition effects are typically targeted by models that do not adjust for total energy intake, while substitution effects are typically targeted by models that do adjust for total energy intake; however, the exact effect being targeted by a particular model will also depend on the other covariates [See box 2 and **Figure 1**]. Where the interest is in an explicit well-defined substitution effect (e.g., fish instead of meat), then only specific modelling strategies will return appropriate estimates. For example, the substitution effect of fish instead of meat can be estimated from a leave-one-out model containing fish (in calories), total energy intake, and all other food groups (in calories) except meat.
- The quality of adjustment for other non-dietary confounding should also be considered and evaluated. Before conducting the review, we recommend constructing directed acyclic graphs (DAGs) for each target causal effect of interest to identify which variables require adjustment to reduce confounding, and which variables should not be adjusted. These DAGs can be used to help identify the most appropriate estimates from each primary study and evaluate the risk of bias for that estimate (29).

**3) For each selected primary study, evaluate the energy adjustment strategy and take careful note of other important modelling decisions**
- Only include estimates from models that use appropriate energy adjustment strategies for your target causal effects of interest. If more than two models target the same causal effect via the same energy adjustment strategy, preference should be given to the model that includes the most confounding variables identified by your DAG but does not include any potential mediators. Where models are missing confounding variables or inappropriately adjust for mediators, consider excluding these estimates outright or exploring their influence through sensitivity analyses.
- Take note of models that adjust for total energy intake in categories. This approach only partially adjusts for total energy intake, leading to estimates that may not be interpretable. Consider excluding these outright or exploring their influence through sensitivity analyses.
- For models where the main exposure is a food (e.g. meat) that contain more than one additional dietary variable (including total energy intake), take note of the (at contain dietary variables in different units (e.g., calories and portions) may not be interpretable. Consider excluding these outright or exploring their influence through sensitivity analyses.
- Take note of models where the exposure has been transformed into a percentage of total energy intake but total energy intake itself is not added as a covariate. Estimates from such models may not be easily interpretable. Consider excluding these outright or exploring their influence through sensitivity analyses.

**4) Pre-specify and clearly report all study details**
- The study aims, target causal effects of interest, appropriate modelling strategies, DAGs, and decisions around other considerations mentioned above should all be clearly pre-specified in the study protocols and clearly reported in the final review articles.
- Where deviations from the protocol are required, e.g., due to unforeseeable issues with the quality or reporting of the primary studies, this should be clearly reported and explained.

**5) Appropriately describe and interpret your results**
- Use clear and appropriate language throughout to describe the study aims, results, and interpretation. Do not refer to unspecified 'associations' or 'relationships'. Language should match the type of causal effect of interest, i.e., addition effects should be described with addition language (e.g., "the effect of consuming an additional 50 g of sugar per day") and substitution effects with substitution language ("the effect of replacing 100 g of red meat per day with 100 g of poultry per day).
- If your study examined more than one causal effect, make sure to interpret each estimate separately and appropriately.
- Do not conduct a meta-analysis when it is not appropriate. Qualitative syntheses should be favoured when the identified studies are too dissimilar to warrant combining into a single value. If a meta-analysis is judged to be appropriate, provide justification for this.

**Conclusion**

Meta-analyses of observational studies of dietary exposures are challenging. Unclear aims and a lack of explicit reasoning about energy adjustment means that many primary studies unwittingly estimate a range of different effects. In an indicative sample, this meta-scientific review found that such issues are not being routinely taken into account by meta-analyses or their constituent studies. Consequently, the pooled estimates relate to ill-defined quantities with unclear interpretations. Both primary studies and meta-analyses need to clearly outline aims and effects of interest. Primary studies should then use energy adjustment strategies that target those effects. Meta-analyses should explicitly consider what effects are being estimated by individual studies and ensure that only similar quantities are pooled. We offer several recommendations to improve the conduct of future meta-analyses, and in turn, hopefully improve the quality of evidence that informs nutritional and public health recommendations.

**Box 1.** Known and new.

| What is already known on this topic |
|---|
| In observational studies of dietary exposures, the effect that is estimated is determined by how the rest of diet, including total energy, is specified in the model. When total energy intake is adjusted for as a covariate or the exposure is expressed as a percentage of total energy intake (e.g., nutrient density), a substitution effect will be estimated; the exact effect being determined by what additional adjustments are made and the distribution of other dietary factors in the study population. When no adjustment is made for total energy intake, then an addition effect will be estimated |
| **What this study adds** |
| Meta-analyses of dietary exposures may not be considering the energy adjustment strategy of the constituent studies, leading to the inappropriate combining of very different effects, including addition effects and a range of different unspecified substitution effects. Consequently, these meta-analyses target ill-defined quantities with unclear interpretations. This study provides a list of recommendations to avoid improve the conduct and interpretability of future meta-analyses of dietary exposures. |

**Box 2 Definition of addition and isocaloric substitution effects, and how they are estimated in observational data**

| |
|---|
| **An addition effect** (also known as a total compositional effect) is the effect of increasing (or decreasing) the consumption of the exposure of interest while keeping the intake of all other foods or macronutrients unchanged. They are estimated by controlling for all other foods or macronutrients, either as separate components or a single summary of remaining energy intake (i.e., total energy intake minus energy from the exposure food or macronutrient of interest). Models that control for other foods or macronutrients in this way are known as energy partition models, where comprehensive energy partition models include all components separately. Subject to assumptions, this approach ensures comparisons are made between groups that differ in terms of a single exposure food or macronutrient of interest. The resulting effect will include both the direct effect of increasing (or decreasing) the exposure food or macronutrient of interest and the indirect effect of simply consuming more (or fewer) calories. |
| **An isocaloric substitution effect** (also known as a relative compositional effect) is the joint effect of increasing one food or macronutrient while decreasing another to maintain the same total energy intake. They are estimated either by subtracting the addition effect of the substituting component from the addition effect of the exposure component or, most commonly, by adjusting for total energy intake. In the latter, the specific substitution being estimated is then determined by which other foods or macronutrients are also adjusted for, where those that are left unadjusted become the substituting components. Models that adjust for total energy intake in this way are known either as standard models or leave-one-out models. Standard models estimate the effect of the exposure component relative to the average effect of all other components. Leave-one-out models estimate the effect of the exposure component relative to the effect of a specific substituting component. To return accurate estimates, standard models and leave-one-out models both require that total energy intake is fully adjusted for and that all components are expressed in calories. Subject to assumptions, these approaches ensure comparisons are made between groups that vary in terms of the exposure, and all substituting foods or macronutrients, but have the same total energy intake. The resulting effect, therefore, does not include the effect of simply consuming more (or less) calories. Estimates of substitution effects are also commonly sought by dividing intake of the exposure food of macronutrient by total energy intake. Models that rescale the exposure in this way are known as nutrient density models, where multivariable nutrient density models additionally adjust for total energy intake. Multivariable nutrient density models provide similar results to standard and leave-one-out models, albeit with rescaled units, but the interpretation of nutrient density models that do not adjust for total calorie intake are more obscure. |

**Table 1.** Descriptive summary on the characteristics of the primary studies included in the meta-analyses on fish and saturated fats.

|  | Saturated Fats n (%) | Fish n (%) |
|---|---|---|
| **Number of primary studies** | 34 | 48 |
| **Linking word in the aim** | 34 | 48 |
|     Association | 17 (50) | 25 (52) |
|     Relationship | 10 (29) | 14 (29) |
|     Influence | 1 (3) | 2 (4) |
|     Effect | 1 (3) | 3 (6) |
|     "Increased risk" | 4 (12) | 2 (4) |
|     "Protective for" | 0 (0) | 1 (2) |
| **Number of exposures** | 45 | 59 |
|     Binary | 1 (2) | 5 (8) |
|     Categorical | 28 (62) | 46 (78) |
|     Continuous | 16 (36) | 8 (14) |
| *Unadjusted models* | 20 (44) | 29 (49) |
| **Number of adjusted models** | 63 | 81 |
| *Units of the exposure* | | |
|     Percentage of calories | 33 (52) | 0 (0) |
|     Mass units (e.g., g/d, g/week) | 30 (48) | 34 (42) |
|     Servings or portions (serv/d, port/wk) | 0 (0) | 28 (35) |
|     Frequency (e.g., occasionally, often, never) | 0 (0) | 19 (23) |
| *Number of models implicitly targeting addition and substitution* | | |
|     *Addition* | 4 (6) | 21 (26) |
|     Not adjusted for total energy intake | 4 (100) | 21 (100) |
|     *Substitution* | 59 (94) | 60 (74) |
|     The model was adjusted for total energy intake as a covariate (standard model, leave-one-out, residual model) | 16 (27) | 54 (90) |
|     The exposure is expressed as a percentage of total calories in the model (nutrient density) | 7 (12) | 1 (2) |
|     Both | 36 (61) | 5 (8) |
| Number of total energy intake adjusted models | 52 (83) | 59 (73) |
|     Coded categorially | 5 (10) | 4 (5) |
|     Justification – Yes *If only listed as potential confounders excluded | 22 (34) | 0 (0) |
| Other dietary variables - Yes | 45 (71) | 54 (90) |
|     Alcohol | 28 (47) | 30 (50) |
|     Protein | 20 (34) | 4 (7) |
|     Fiber | 17 (29) | 14 (23) |
|     Vegetables | 13 (22) | 24 (40) |
|     Fruits | 10 (17) | 24 (40) |
|     Dietary cholesterol | 8 (14) | 2 (3) |
| Carbohydrates | 6 (10) | 5 (8) |
| Vitamin E and C | 5 (8) | 2 (3) |
| Polyunsaturated fats | 5 (8) | 8 (14) |
| Saturated fats | NA | 14 (23) |
| Red meat | 0 | 6 (10) |
| Mediterranean diet score | 0 | 4 (7) |
| Homogeneity units among substitutive models | | |
| Yes | 6 (10) | 7 (12) |
| No | 49 (83) | 50 (83) |
| NA (Only one dietary variable) | 4 (7) | 3 (5) |

**Table 2. Checklist for meta-analyses in nutritional epidemiology focusing on foods and macronutrients.**

| Section and topic | Checklist item | Page-line number |
|---|---|---|
| **Aims** | | |
| Objective | 1. State your target causal effects of interest, clearly indicating if they are addition or substitution effects.<br>    a. If interested in substitution effects, state which substituting components are of interest<br>2. Formulate the target causal effects using an explicit framework such as PICOT. | |
| **Methods** | | |
| | 3. For each target causal effect of interest, identify all appropriate energy adjustment strategies.<br>4. For each target causal effect of interest, construct a directed acyclic graph to identify appropriate and necessary variables for adjustment<br>5. For each estimate in each study, extract all information relevant to the quantity being estimated<br>    a. Note what dietary variables have been adjusted for<br>    b. Note what non-dietary variables have been adjusted for and whether these are confounders or mediators<br>    c. For models that contain total energy intake, note whether the variable has been categorised<br>    d. For models that include more than one dietary variable, take note of models that contain dietary variables in different units<br>    e. Take note when the exposure has been transformed into a percentage of total energy intake where total energy intake is not also a covariate.<br>    f. Where the exposure has been categorised, note the mean and range of the categories being compared<br>    g. Note whether the causal effect of interest is possible in the population being studied.<br>6. Pre-specify your preferred approach to key methodological weaknesses (such as categorisation of total energy intake), i.e. whether such studies will be excluded outright or subject to sensitivity analyses. | |
| **Results** | | |
| | 7. Only pool and report estimates for similar target quantities<br>    a. Conduct qualitative syntheses when the identified studies are too dissimilar to warrant combining<br>8. If the study examines more than one causal effect, interpret each estimate distinctively | |

| **Discussion** | |
| --- | --- |
| | 9. Use clear and appropriate language throughout to describes aims, results, and interpretations<br>    a. Do not refer to unspecified 'associations' or 'relationships'.<br>    b. Describe addition or substitution effects with appropriate language.<br>10. Provide a detailed discussion interpretating the estimates and sources of variation that were not quantitatively explored that you may have not described in the results. |

**Figure 1.** Common causal effects targeted by different energy adjustment strategies in nutritional epidemiology studies.

| Reference diet | Comparison diet | Description of effect | Model |
|---|---|---|---|
| **Addition effects** | | | |
| | | Average effect of increasing intake of the exposure component (yellow) plus all typical changes in the diet that would accompany increasing intake of the exposure component (red, blue, and green). | **Outcome ~ exposure** + confounders |
| | | Average effect of increasing intake of the exposure component (yellow) assuming no other changes to the diet. | **Outcome ~ exposure + component_1 + component_2 + component_3** + confounders |
| | | Average effect of increasing intake of the exposure component (yellow) assuming no changes in the calorie intake from all other component but including any typical changes in the proportions of all other components (i.e. red, blue, and green). | Outcome ~ **exposure + remaining_energy** * + confounders |
| **Substitution effects** | | | |
| | | Average effect of increasing intake of the exposure component (yellow) while decreasing intake of all other components (red, blue, and green) in an unspecified manner to maintain the same total calorie intake. | Outcome ~ **exposure + total_energy** + confounders |
| | | Average effect of increasing intake of the exposure component (yellow) while maintaining the same intake of one component (green) and decreasing intake of the other components (red and blue) in an unspecified manner to maintain the same total calorie intake. | Outcome ~ **exposure + total_energy + component_3** + confounders |
| | | Average effect of increasing intake of the exposure component (yellow) while decreasing intake of a specified other component (red) and maintaining the same intake of all remaining components (green and blue) to maintain the same total calorie intake. | Outcome ~ **exposure + total_energy + component_2 + component_3** + confounders |

**Data availability statement**
The full dataset of extracted data, discrepancies and quality check of randomly selected studies is publicly available on OSF (7).

**Patient and Public involvement statement**
Patients or members of the public were not involved in the design, conduct, reporting, or dissemination of this research. This is because the manuscript focuses on improving research practices within nutritional science, with researchers as the primary audience. To enhance relevance and applicability for the target community, we engaged co-authors from a range of disciplinary and geographical backgrounds.

**Funding**
NO was awarded a "UniBE Short Travel Grants for (Post)Docs" and a Postdoc.Mobility grant by the Swiss National Science Foundation (n 230635). DBI acknowledges support from the Independent Research Fund Denmark (1057-00016B) and the Danish Diabetes Association. No other authors received funding for this study.

**Conflicts of interest**
PWGT is a director of Causal Thinking Ltd, which provides causal inference research and training. He may therefore benefit from any study that demonstrates the value of causal inference methods.
All other authors declare no conflicts of interest related to this manuscript.

**Author contributions (Credit statement)**
NO, PWGT, GDT conceptualized the project and designed the study. All co-authors equally contributed to the data extraction. NO did the formal analysis and NO and PWGT wrote the original draft. All authors contributed to critically reviewing and editing the draft and approved the final manuscript before submission. GDT was responsible for the supervision of the project.